\documentclass[preprint]{aastex}

\slugcomment{To appear in Ap. J. Lett.}

\shorttitle{VERITAS Upper Limit on the Radio Galaxy NGC~1275}
\shortauthors{Acciari et al.}

\begin{document}


\title{VERITAS Upper Limit on the VHE Emission from the Radio Galaxy NGC~1275}


\author{
V.~A.~Acciari\altaffilmark{1,22},
E.~Aliu\altaffilmark{2},
T.~Arlen\altaffilmark{3},
T.~Aune\altaffilmark{4},
M.~Bautista\altaffilmark{5},
M.~Beilicke\altaffilmark{6},
W.~Benbow\altaffilmark{1},
D.~Boltuch\altaffilmark{2},
S.~M.~Bradbury\altaffilmark{7},
J.~H.~Buckley\altaffilmark{6},
V.~Bugaev\altaffilmark{6},
K.~Byrum\altaffilmark{8},
A.~Cannon\altaffilmark{9},
O.~Celik\altaffilmark{3},
A.~Cesarini\altaffilmark{10},
L.~Ciupik\altaffilmark{11},
P.~Cogan\altaffilmark{5},
W.~Cui\altaffilmark{12},
R.~Dickherber\altaffilmark{6},
C.~Duke\altaffilmark{13},
S.~J.~Fegan\altaffilmark{3},
J.~P.~Finley\altaffilmark{12},
P.~Fortin\altaffilmark{14},
L.~Fortson\altaffilmark{11},
A.~Furniss\altaffilmark{4},
N.~Galante\altaffilmark{1,*},
D.~Gall\altaffilmark{12},
K.~Gibbs\altaffilmark{1},
G.~H.~Gillanders\altaffilmark{10},
S.~Godambe\altaffilmark{15},
J.~Grube\altaffilmark{9},
R.~Guenette\altaffilmark{5},
G.~Gyuk\altaffilmark{11},
D.~Hanna\altaffilmark{5},
J.~Holder\altaffilmark{2},
D.~Horan\altaffilmark{16},
C.~M.~Hui\altaffilmark{15},
T.~B.~Humensky\altaffilmark{24},
A.~Imran\altaffilmark{17},
P.~Kaaret\altaffilmark{18},
N.~Karlsson\altaffilmark{11},
M.~Kertzman\altaffilmark{19},
D.~Kieda\altaffilmark{15},
A.~Konopelko\altaffilmark{20},
H.~Krawczynski\altaffilmark{6},
F.~Krennrich\altaffilmark{17},
M.~J.~Lang\altaffilmark{10},
S.~LeBohec\altaffilmark{15},
G.~Maier\altaffilmark{5},
A.~McCann\altaffilmark{5},
M.~McCutcheon\altaffilmark{5},
J.~Millis\altaffilmark{21},
P.~Moriarty\altaffilmark{22},
R.~Mukherjee\altaffilmark{14},
R.~A.~Ong\altaffilmark{3},
A.~N.~Otte\altaffilmark{4},
D.~Pandel\altaffilmark{18},
J.~S.~Perkins\altaffilmark{1},
M.~Pohl\altaffilmark{17},
J.~Quinn\altaffilmark{9},
K.~Ragan\altaffilmark{5},
P.~T.~Reynolds\altaffilmark{23},
E.~Roache\altaffilmark{1},
H.~J.~Rose\altaffilmark{7},
M.~Schroedter\altaffilmark{17},
G.~H.~Sembroski\altaffilmark{12},
A.~W.~Smith\altaffilmark{8},
D.~Steele\altaffilmark{11},
S.~P.~Swordy\altaffilmark{24},
M.~Theiling\altaffilmark{1,25},
J.~A.~Toner\altaffilmark{10},
A.~Varlotta\altaffilmark{12},
V.~V.~Vassiliev\altaffilmark{3},
S.~Vincent\altaffilmark{15},
R.~G.~Wagner\altaffilmark{8},
S.~P.~Wakely\altaffilmark{24},
J.~E.~Ward\altaffilmark{9},
T.~C.~Weekes\altaffilmark{1},
A.~Weinstein\altaffilmark{3},
T.~Weisgarber\altaffilmark{24},
D.~A.~Williams\altaffilmark{4},
S.~Wissel\altaffilmark{24},
M.~Wood\altaffilmark{3},
B.~Zitzer\altaffilmark{12}
}

\author{
J.~Kataoka\altaffilmark{26},
E.~Cavazzuti\altaffilmark{27},
C.~C.~Cheung\altaffilmark{28},
B.~Lott\altaffilmark{29,30},
D.~J.~Thompson\altaffilmark{28},
G.~Tosti\altaffilmark{31,32}
}

\altaffiltext{*}{Corresponding Author: ngalante@cfa.harvard.edu}
\altaffiltext{1}{Fred Lawrence Whipple Observatory, Harvard-Smithsonian Center for Astrophysics, Amado, AZ 85645, USA}
\altaffiltext{2}{Department of Physics and Astronomy and the Bartol Research Institute, University of Delaware, Newark, DE 19716, USA}
\altaffiltext{3}{Department of Physics and Astronomy, University of California, Los Angeles, CA 90095, USA}
\altaffiltext{4}{Santa Cruz Institute for Particle Physics and Department of Physics, University of California, Santa Cruz, CA 95064, USA}
\altaffiltext{5}{Physics Department, McGill University, Montreal, QC H3A 2T8, Canada}
\altaffiltext{6}{Department of Physics, Washington University, St. Louis, MO 63130, USA}
\altaffiltext{7}{School of Physics and Astronomy, University of Leeds, Leeds, LS2 9JT, UK}
\altaffiltext{8}{Argonne National Laboratory, 9700 S. Cass Avenue, Argonne, IL 60439, USA}
\altaffiltext{9}{School of Physics, University College Dublin, Belfield, Dublin 4, Ireland}
\altaffiltext{10}{School of Physics, National University of Ireland, Galway, Ireland}
\altaffiltext{11}{Astronomy Department, Adler Planetarium and Astronomy Museum, Chicago, IL 60605, USA}
\altaffiltext{12}{Department of Physics, Purdue University, West Lafayette, IN 47907, USA }
\altaffiltext{13}{Department of Physics, Grinnell College, Grinnell, IA 50112-1690, USA}
\altaffiltext{14}{Department of Physics and Astronomy, Barnard College, Columbia University, NY 10027, USA}
\altaffiltext{15}{Department of Physics and Astronomy, University of Utah, Salt Lake City, UT 84112, USA}
\altaffiltext{16}{Laboratoire Leprince-Ringuet, Ecole Polytechnique, CNRS/IN2P3, F-91128 Palaiseau, France}
\altaffiltext{17}{Department of Physics and Astronomy, Iowa State University, Ames, IA 50011, USA}
\altaffiltext{18}{Department of Physics and Astronomy, University of Iowa, Van Allen Hall, Iowa City, IA 52242, USA}
\altaffiltext{19}{Department of Physics and Astronomy, DePauw University, Greencastle, IN 46135-0037, USA}
\altaffiltext{20}{Department of Physics, Pittsburg State University, 1701 South Broadway, Pittsburg, KS 66762, USA}
\altaffiltext{21}{Department of Physics, Anderson University, 1100 East 5th Street, Anderson, IN 46012}
\altaffiltext{22}{Department of Life and Physical Sciences, Galway-Mayo Institute of Technology, Dublin Road, Galway, Ireland}
\altaffiltext{23}{Department of Applied Physics and Instrumentation, Cork Institute of Technology, Bishopstown, Cork, Ireland}
\altaffiltext{24}{Enrico Fermi Institute, University of Chicago, Chicago, IL 60637, USA}
\altaffiltext{25}{Department of Physics and Astronomy, Clemson University, Clemson, SC 29634, USA}
\altaffiltext{26}{Research Institute for Science and Engineering, Waseda University, 3-4-1, Okubo, Shinjuku, Tokyo, 169-8555, Japan}
\altaffiltext{27}{Agenzia Spaziale Italiana (ASI) Science Data Center, I-00044 Frascati (Roma), Italy}
\altaffiltext{28}{NASA Goddard Space Flight Center, Greenbelt, MD 20771}
\altaffiltext{29}{CNRS/IN2P3, Centre dÕ\'Etudes Nucl\'eaires Bordeaux Gradignan, UMR 5797, Gradignan, 33175, France}
\altaffiltext{30}{Universit\'e de Bordeaux, Centre dÕ\'Etudes Nucl\'eaires Bordeaux Gradignan, UMR 5797, Gradignan, 33175, France}
\altaffiltext{31}{Istituto Nazionale di Fisica Nucleare, Sezione di Perugia, I-06123 Perugia, Italy}
\altaffiltext{32}{Dipartimento di Fisica, Universit\`a degli Studi di Perugia, I-06123 Perugia, Italy}

\begin{abstract}
The recent detection by the \emph{Fermi} $\gamma$-ray space telescope 
of high-energy $\gamma$-rays from the radio galaxy NGC~1275 makes the observation of the 
very high energy (VHE: $E>100$~GeV) 
part of its broadband spectrum particularly interesting, especially for the understanding of 
active galactic nuclei (AGN) with
misaligned multi-structured jets. The radio galaxy NGC~1275 was recently observed by VERITAS at energies above 100~GeV
for about 8~hours.
No VHE $\gamma$-ray emission was detected by VERITAS from NGC~1275.
A 99\% confidence level  upper limit of 2.1\% of the Crab Nebula flux level is obtained at the decorrelation energy of approximately 
340~GeV, corresponding to 19\% of the power-law extrapolation of the \emph{Fermi} 
Large Area Telescope (LAT) result.
\end{abstract}





\section{Introduction}

The search for $\gamma$-rays from radio galaxies is  important for
the understanding of the dynamics and structure of jets in active galactic nuclei (AGN).
Radio galaxies are the only non-blazar extragalactic objects detected in the
very high energy range so far. Even though radio galaxies contain jets and are
considered AGN, their jet is not oriented toward the observer
and therefore the radiation produced by the jet is not Doppler-boosted towards
higher energies and luminosities, making them more challenging to detect in the 
very high energy (VHE: $E>100$~GeV) regime.
The discovery of VHE  $\gamma$-rays from the radio galaxy M~87 by the HEGRA
collaboration~\citep{goetting2003}, detected later by VERITAS~\citep{Acciari2008a},
and from NGC~5128 (Centaurus~A) by the HESS collaboration~\citep{Raue2009} has shown that
non-blazar AGN can produce very energetic photons from non-thermal processes.

Radio galaxies are classified into two main
families based on the morphology of their radio emission~\citep{FanaroffRiley},
whether it is core dominated (FR~I) or lobe dominated (FR~II).
The differences between these two families can be found in the radio energetics and
in the discrete spectral properties~\citep{Zirbel1995}. The large number of features that
FR~I radio galaxies share with BL Lac type blazars suggests a possible unification between
the two sub-classes of AGN, in which FR I radio galaxies are BL Lac objects observed at larger jet viewing
angles~\citep{UrryPadovani}.

Evidence for synchrotron emission in radio to X-ray energies from both the extended structures and 
the core is well explained by relativistic particles moving in a beamed
relativistic jet~\citep{Ghisellini1993}. 
A commonly considered mechanism for HE-VHE (HE: high energy, 100~MeV$<E<$100~GeV) radiation is the synchrotron-self-Compton (SSC) 
process~\citep{Jones1974}, where the optical and UV synchrotron photons are up-scattered by the same 
relativistic electrons in the jet. Predictions concerning the inverse Compton (IC) component 
have long been established for the $\gamma$-ray 
emission~\citep{BloomMarscher1996} and frequency-dependent 
variability~\citep{Ghisellini1989}. Besides leptonic scenarios, several models also consider a hadronic origin for 
non-thermal emission in jets. Accelerated protons can initiate electromagnetic cascades or 
photomeson processes~\citep{Mannheim1993}, or directly emit synchrotron radiation \citep{Aharonian2002, Reimer2004}
and produce $\gamma$-rays through collisions with ambient gas \citep{Beall1999, Pohl2000}.

Modelling the blazar jet emission with a homogeneous SSC mechanism may imply particularly
high Lorentz factors, $\Gamma \gtrsim 50$, with consequent high Doppler factors and small beaming angles $\theta \simeq 1^\circ$
\citep{Kraw2002}. Such a small beaming angle is in conflict with the unification scheme according to which FR~I radio galaxies
and BL~Lac objects are the same kind of object observed at different viewing angles. Moreover,
these high values for the Doppler factor are in disagreement with the small apparent velocities observed
in the sub-parsec region of the TeV BL Lac objects Mrk~421 and Mrk~501 \citep{Marscher1999}.
These considerations suggest a more complicated geometry, for example 
a decelerating flow in the jet with a consequent gradient in the Lorentz
factor of the accelerated particles and a smaller average $\Gamma$ \citep{Georganopoulos2003}.
As a result of this gradient, the fast upstream particles interact with the downstream 
seed photons with an amplified energy density, because of the Doppler boost due to the relative Lorentz factor
$\Gamma_\mathrm{rel}$. The IC process then requires less extreme values for the
Lorentz factor and allows larger values for the beaming angle.
In a similar way, a jet spine-sheath structure consisting of a faster internal spine 
surrounded by a slower layer has been also suggested for the broadband non-thermal emission of VHE BL Lac 
objects~\citep{Ghisellini2005}. This model is supported by radio maps which show a limb-brightened 
morphology ~\citep{Giroletti2004} and may explain the HE-VHE emission observed in radio galaxies at larger angles              
($\theta_\mathrm{layer} = 1/\Gamma_\mathrm{layer} \sim 20^\circ$). 
Observation of the VHE component from radio galaxies
is therefore significant for the multi-structured jet modeling, as it can be related to the external lower
structure of the jet itself.

NGC~1275 (Perseus~A, 3C~84) is a radio galaxy located in the center of the Perseus cluster and is one of the most
unusual early-type galaxies in the nearby universe ($z=0.018$). Its radio emission is core dominated, but it also has strong
emission lines. In addition, the emission line system shows a double structure, 
corresponding to both a high-velocity and a low-velocity 
system. The puzzling nature of NGC~1275 makes it difficult to definitively classify it
in a standard AGN sub-class. Its spectral energy distribution (SED) extends from radio to HE $\gamma$-rays. 
A source coincident with NGC~1275 at high confidence was
recently detected in HE $\gamma$-rays by the \emph{Fermi} $\gamma$-ray space telescope
\citep{Abdo2009a}. The time-average spectrum observed
between August and December 2008 is described by a power law in the energy range 
from 100~MeV to 25~GeV :

\begin{equation}\label{eq:1}
\frac{\mathrm{d}N}{\mathrm{d}E} = (2.45\pm 0.26) \times 10^{-9} \left(\frac{E}{100\,\mathrm{MeV}}\right)^{-2.17\pm 0.04} \qquad\mathrm{cm}^{-2}\; \mathrm{s}^{-1}\; \mathrm{MeV}^{-1}
\end{equation}

\noindent A radio luminosity similar to the typical values ($L_\mathrm{r} = 10^{42}$~\mbox{[erg~s$^{-1}$]}) 
of the BL~Lacs in the Large Area Telescope (LAT) bright AGN sample 
has been also reported \citep{Abdo2009b},
but no VHE detection above 100~GeV has been reported so far.

The broad-band SED of NGC~1275 shows similarities with low-frequency-peaked BL Lac (LBL) objects.
VHE $\gamma$-ray emission from LBLs is already established, namely from BL~Lac~\citep{Albert2007} and
from the two  intermediate-frequency-peaked BL Lac (IBL) 
3C~66A~\citep{Acciari2009} and W~Comae~\citep{Acciari2008b}. NGC~1275 is thus a potential VHE source.

\section{VERITAS Observation}


The VERITAS detector is an array of four 12-m diameter imaging
atmospheric-Cherenkov telescopes located in southern Arizona
\citep{Weekes:2002pi}.  Designed to detect emission from astrophysical
objects in the energy range from 100~GeV to greater than 30~TeV,
VERITAS has an energy resolution of $\sim$15\% and an angular
resolution (68\% containment) of $\sim$$0.1^\circ$ per event at 1~TeV.  A source
with a flux of 1\% of the Crab Nebula flux is detected in less than 50~hours
of observations, while a \mbox{5\% Crab Nebula} flux source is detected in
2.5~hours.  The field of view of the VERITAS telescopes is
$3.5^\circ$.  For more details on the VERITAS instrument and the imaging atmospheric-Cherenkov
technique, see \citet{Holder:2008}.

The \emph{Fermi} $\gamma$-ray space telescope reported to its multiwavelength partners
the detection of NGC~1275 and a measurement of a hard $\gamma$-ray spectrum in the HE range in Fall 2008.
VERITAS observed the core region of NGC~1275 for about 11~hours between January 15$^\mathrm{th}$ and 
February 26$^\mathrm{th}$ 2009. Data taken with poor weather conditions or technical problems are excluded,
resulting in 7.8~hours of live time.
The average zenith angle of the VERITAS observation is $30^\circ$. All data were taken
in ``wobble'' mode \citep{fomin1994} where the telescopes are
pointed away from the source by $0.5^\circ$ 
North/South/East/West to allow for
simultaneous background estimation using events from the same field of view.

\section{Analysis Methods}
   
Prior to event selection and background subtraction, all shower images
are calibrated and cleaned as described in \cite{Cogan:2006} and
\cite{Daniel:2007kx}.  Several noise-reducing event-selection cuts are
made at this point, including rejecting those events where only the
two closest-spaced telescopes participated in the trigger. Following
the calibration and cleaning of the data, the events are parametrized
using a moment analysis \citep{Hillas:1985ta}.  From this moment
analysis, scaled parameters are calculated and used for the
selection of the $\gamma$-ray-like events \citep{Aharonian:1997rm,Krawczynski:2006ts}. The event
selection cuts are optimized {\it a priori} on a weak Crab-like source (photon index $\alpha=2.5$ and 3\%~Crab Nebula flux level). 
These selection criteria are termed the ``standard cuts''. 
As the VHE spectrum of NGC~1275 is unknown, the possibility
that it could be considerably softer (photon index $\alpha>4$) 
due to internal (slower bulk motion) or external (dust absorption) factors has been considered. A set of
modified ``soft cuts'' are also applied by increasing the $\theta^2$ cut
(the square of the angular distance of the nominal position of NGC~1275 from the reconstructed 
air shower direction) and decreasing the size cut (the minimum
number of photo-electrons required in an image). Table~\ref{tab:result} shows the
set of cuts used in this work.  Unless stated otherwise, the
``standard cuts'' are used to generate the results presented in this
paper.

\begin{table}
\begin{center}
  \caption{Selection cuts applied to the VERITAS data for NGC~1275 and the analysis results. 
  For each of the two sets of cuts, the nine columns show: 
  the $\theta^2$ cut; the {\it size} cut in digital counts \mbox{(1 DC $\simeq$ 4.5 photoelectrons)}; the analysis energy threshold; the decorrelation energy; the total number
  of excess events; the corresponding significance according to Formula~17 in \cite{Li:1983pb}; the differential
  upper limit at the decorrelation energy;  the same upper limit in Crab Nebula units; the integral upper limit
  above the analysis energy threshold.
  A total of 9 and 5 background regions have been used for the standard and soft cuts analysis respectively.
  A power-law photon index of $-2.5$ has been assumed in the flux upper limit calculation.
  It is important to note that the upper limits for the standard and soft cuts are highly correlated.}
    \label{tab:result}
\begin{tabular}{cccccccccc}
\tableline\tableline
Cuts & $\theta^2$ & Size & $E_\mathrm{th}$ & $E_\mathrm{decorr}$ & N$_\mathrm{ex}$ &  N$_{\sigma}$ & \multicolumn{3}{c}{99\% C. L. upper limit}\\
     & [deg$^2$] & [DC] & [GeV] & [GeV]  & && [cm$^{-2}$ TeV$^{-1}$ s$^{-1}$] & C.U.  & [cm$^{-2}$ s$^{-1}$]\\
\tableline
Soft & $< 0.040$ & $> 200$ & 126 & 235 & 10.4 & 0.7 &$3.19\times 10^{-11}$ & 0.028 & $1.27\times 10^{-11}$   \\
Std. & $< 0.017$ & $> 400$ & 188 & 338 & -13.0 & -0.3 &$9.43\times 10^{-12}$ & 0.021 & $5.11\times 10^{-12}$   \\
\tableline
\end{tabular}
\end{center}
\end{table}

The reflected-region model \citep{Berge:2007ud} is used for background
subtraction. The total number of events in the on-source region is
then compared to the total number of events in the more numerous
off-source regions, scaled by the ratio of the solid angles, to produce a final excess.

\section{Fermi Observation}

The published result from \emph{Fermi} observation does not overlap with the VERITAS observation campaign.
Additional data taken during the period from January 15$^\mathrm{th}$ to February 26$^\mathrm{th}$ 2009
have been therefore used in this work. A zenith angle cut at 105$^\circ$ 
has been applied to eliminate photons from the Earth limb.

The analysis follows the same procedure described in detail in~\cite{Abdo2009a}. A lower energy cut of 200~MeV
has been applied to avoid strong systematic uncertainties of the lowest energy events.
Science Tools version v9r15p2 and IRFs (Instrumental Response Functions) P6\_V3\_Diffuse (a model of the spatial 
distribution of photon events calibrated pre-launch) were used throughout this paper.
The \emph{Fermi}-LAT data reveal a signal in the energy range between 200~MeV and 6.4~GeV
of 14.0~$\sigma$ significance at the source location.

\section{Result and discussion}

\begin{figure}
\epsscale{.90}
\plotone{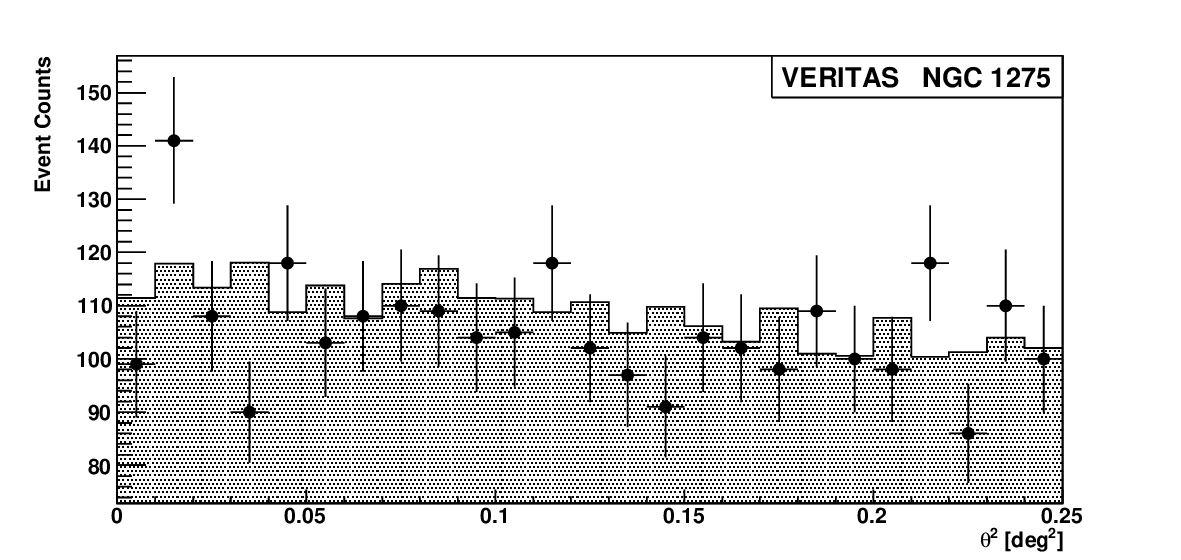}
\caption{$\theta^2$ plot for VERITAS observation of NGC~1275. The points with error bars represent the
ON-source data sample and the filled area is the background.}
\label{fig:theta2}
\end{figure}

\begin{figure}
\epsscale{.90}
\plotone{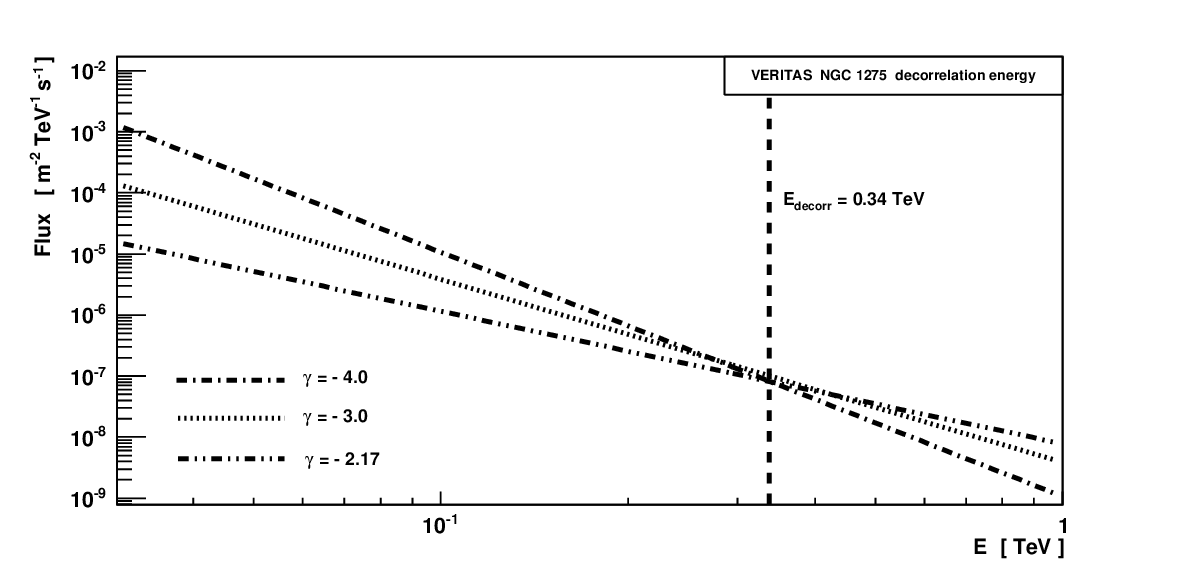}
\caption{Decorrelation energy for the standard-cuts analysis on VERITAS data.
Three different photon indices, $-2.17$, $-3.0$ and $-4.0$, are assumed for the power law used to calculate the flux upper limit.
The central value of the interval where the three normalized power laws intersect is taken as decorrelation energy.
The dependency of the differential flux upper limit on the assumed photon index is strongly reduced at the decorrelation energy.
Assuming a photon index $-2.5$, the effect of the photon index change within the values here considered at the decorrelation
energy is about 15\%.}
\label{fig:decorr}
\end{figure}

No significant excess is detected by VERITAS in the NGC~1275 data sample. 
Figure~\ref{fig:theta2} shows the $\theta^2$ plot resulting from the
standard cuts analysis. A 99\% confidence level upper limit is calculated for both sets
of cuts using the method described by~\cite{Rolke2005} and assuming a Gaussian background.
The results are reported in Table~\ref{tab:result}
where the differential flux upper limit at the decorrelation energy is shown.
We define decorrelation energy the energy that minimizes the dependency of the differential upper limit
on the assumed photon index. Three different photon indices are considered and three
``upper limit functions" are hence calculated. The central value of the interval where they intersect is
taken as decorrelation energy as shown in figure~\ref{fig:decorr}.
An integral flux upper limit above the analysis energy threshold 
(the energy corresponding to the maximum of the product function of observed spectrum
with the collection area, 188~GeV for standard cuts) is also shown.
The soft-cuts analysis has the effect to lower the analysis energy threshold but also to increase
the amount of residual background events. Since the decorrelation energy is in both analysises
above the standard-cuts analysis energy threshold (see table~\ref{tab:result}), we discuss hereafter only the
result from the standard-cuts analysis as it provides a stronger background reduction.

The energy-binned analysis of the \emph{Fermi} data taken over the VERITAS observation campaign
shows a spectrum that can be fitted with the following power law from 200~MeV to 6.4~GeV

\begin{equation}
\frac{\mathrm{d}N}{\mathrm{d}E} = (1.74\pm 0.45) \times 10^{-9} \left(\frac{E}{100\,\mathrm{MeV}}\right)^{-2.15\pm 0.11} \qquad\mathrm{cm}^{-2}\; \mathrm{s}^{-1}\; \mathrm{MeV}^{-1}
\end{equation}

\noindent There are some differences from the spectrum presented in~\cite{Abdo2009a} and summarized
in formula~(\ref{eq:1}). The photon index is compatible within the error but the integral flux between
200~MeV and 6.4~GeV is a factor of 1.37 lower, $F^{09}_\mathrm{int} = (6.7\pm0.88)\times10^{-8}$~\mbox{cm$^{-2}$ s$^{-1}$},
compatible with the 2008 integral flux within 2.8 standard deviations ($F^{08}_\mathrm{int} = (9.19\pm0.50)\times10^{-8}$~\mbox{cm$^{-2}$ s$^{-1}$}).
The lower flux comfirms the hint of decreasing trend of the flux that was discussed in~\cite{Abdo2009a},
although the difference is statistically weak.
In addition, the 2009 spectrum does not extend
beyond $\sim$6~GeV, because of the overall lower flux and the limited observation time
that both reduce the statistical significance of the events at higher energy.

\begin{figure}
\plotone{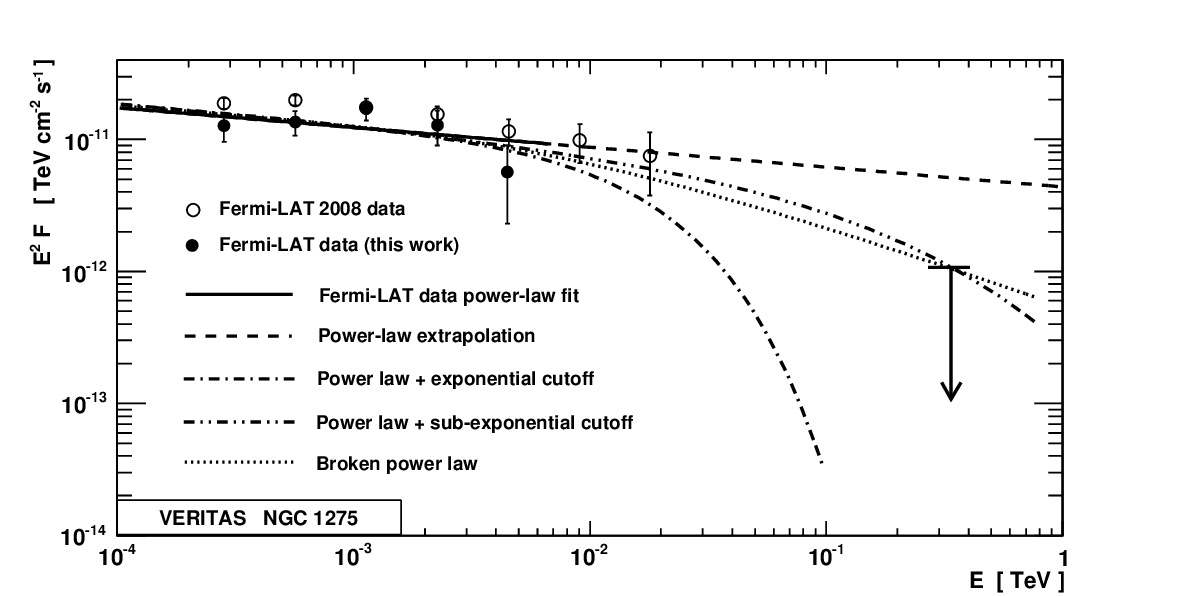}
\caption{NGC~1275 spectrum and the VERITAS upper limit on the differential flux at 
the decorrelation energy 338~GeV (standard cuts). The solid circles with error bars are the measurement
by the \emph{Fermi} $\gamma$-ray space telescope during the VERITAS observation campaign. 
Empty circles with error bars are the measurement presented in~\cite{Abdo2009a} from the energy-binned analysis. 
The solid line is the power-law fit to the \emph{Fermi} data. 
The dashed line is the extrapolation of the power-law. The dotted-dashed line is the fit of
a power law with an exponential cutoff at 18~GeV. The double-dotted dashed line
is the fit of a power law with a sub-exponential cutoff at 120~GeV and the dotted line is
the smooth broken power law fit of a break energy at 16~GeV. All fits are done on the \emph{Fermi} data
analyzed in this work (solid circles).}
\label{fig:ul}
\end{figure}

Figure~\ref{fig:ul} shows the VERITAS differential flux upper limit,
calculated at the decorrelation energy of the standard-cuts analysis, together with the \emph{Fermi} spectra.
The lower statistical significance of the overall signal in the \emph{Fermi} data simultaneous with the VERITAS
observation, compared to the measurement in~\cite{Abdo2009a}, is visible in the larger error bars of the spectral points.
The VERITAS upper limit is a factor of five below the extrapolation of the power-law 
fit between 200~MeV and 6.4~GeV of the \emph{Fermi} data taken during the VERITAS observation campaign.
Considering the proximity of the NGC~1275 galaxy ($z=0.018$), negligible absorption by the
interaction of VHE $\gamma$-rays on the extra-galactic background light (EBL) is expected.
Given three degrees of freedom for the data sample, 
the VERITAS upper limit implies a power-law fit with a normalized $\chi^2/3$ value of 2.97, corresponding
to a probability of $P(\chi^2) = 0.03$. The large value of the $\chi^2$ probability is given by the large error
bars of the spectral points due to the relatively short observation time and to the lower flux 
than the measurement in~\cite{Abdo2009a}, that both reduce the overall statistics of events especially at higher energy.
Since the photon index does not change significantly and the extrapolated flux is still
larger than the VERITAS flux upper limit, the same statistical test has been done on the
spectrum in~\cite{Abdo2009a} after applying a scaling factor of 1.35 to the spectrum itself. The result
gives a normalized $\chi^2/5$ value of 4.57 with a probability $P(\chi^2) = 3.6\times 10^{-4}$.
We conclude that the upper limit is incompatible with an extrapolation of the power law from the \emph{Fermi} data. 


A deviation from the pure power-law spectrum is suggested.
The \emph{Fermi} data analyzed in this work are fitted using a power law with
the addition of an exponential cutoff, imposing the constraint given by the upper limit.
The VHE result is compatible with this function if the cut-off energy is \mbox{$E_\mathrm{cut}\approx 18$ GeV}.
A normalized $\chi^2/3 = 1.44$ is obtained for the fit.
 A power law with such an exponential cutoff in the IC component may
arise from an electron energy distribution that follows a power law with a rather sharp cutoff. 

Alternatively, if the energy distribution of the accelerated electrons
in the jet follows a power law with an exponential cutoff in the form:

\begin{equation}\label{eq:electrons1}
N(\gamma) \varpropto \gamma^{-p}\, e^{-\gamma/\gamma_\mathrm{c}}
\end{equation}

\noindent where $p$ is the electron spectral index, $\gamma$ is the particle Lorentz factor and $\gamma_\mathrm{c}$ is the
Lorentz factor at the cut-off energy, the emitted synchrotron radiation flux  and IC component, as a function of energy $\phi(E)$,
would have a sub-exponential cutoff in the form:

\begin{equation}\label{eq:expcutoff}
\phi(E) \varpropto E^{-(p-1)/2}\, e^{-\sqrt{E/E_\mathrm{cut}}}
\end{equation}

This sub-exponential cutoff is a reasonable approximation if the IC scattering proceeds in the Thomson regime.
Fitting the \emph{Fermi}-LAT data and including the 
VERITAS upper limit gives a cut-off energy $E_\mathrm{cut}\approx 120$~GeV
and $\chi^2/3= 1.56$. Probabilities of the $\chi^2$ quantile with five degrees of freedom
are 0.23 and 0.20 for the exponential and sub-exponential cut-off result, respectively. The fit with a sub-exponential
cutoff gives a smaller normalized $\chi^2$ value than the fit with the sub-exponential cutoff,
but the difference is not significant. Both assumptions are therefore
compatible with \emph{Fermi}-LAT and VERITAS measurements.

Another possibility is that the measured power-law steepens to a softer spectral index. This behaviour
arises if the accelerated electrons below a certain energy
escape from their environment faster than the typical
synchrotron radiation time. Assuming an electron injection in the jet following a power law $Q(\gamma)\varpropto \gamma^{-p}$ 
and a constant and energy-independent leakage of particles per unit time, the stationary electron energy distribution
may be approximated with the form:

\begin{equation}
N(\gamma) \varpropto \gamma^{-p} \; \left( 1-e^{-\gamma_\mathrm{b}/\gamma} \right)
\end{equation}

\noindent where $\gamma_\mathrm{b}$ is the energy at which the energy loss due to the leakage
is equal to the synchrotron radiating energy loss. The IC component of the SED would assume
the following form:

\begin{equation}\label{eq:ICbroken}
P(\nu) = \nu F(\nu) \varpropto \nu^{(1-p)/{2}} \, \left( 1-e^{-\sqrt{\nu_\mathrm{b}/{\nu}}} \right)
\end{equation}

\noindent where $\nu_\mathrm{b}$ is the break frequency in the IC component. 
For $\nu \ll \nu_\mathrm{b}$, equation~(\ref{eq:ICbroken}) asymptotically follows the power law
$\nu^{({1-p})/{2}}$, and for $\nu \gg \nu_\mathrm{b}$ the asymptotic limit becomes
$\nu^{-{p}/{2}}$. Equation~(\ref{eq:ICbroken}) thus represents a smooth transition from a power-law
regime below the break frequency $\nu_\mathrm{b}$ to a steeper power-law regime above the break frequency.
In this scenario the upper limit obtained by the VERITAS measurement implies a break energy
$E_\mathrm{b} \approx 16$~GeV. The normalized $\chi^2$ obtained from the fit with function~(\ref{eq:ICbroken})
is $\chi^2/3 = 1.41$. The probability of the $\chi^2$ quantile with 3 degrees of freedom is 0.24, 
similar to the probabilities obtained in the exponential- and sub-exponential-cutoff scenarios
here considered.

It must be noticed that the result on the cutoff or break energy does not change significantly in the soft-cuts analysis
since the corresponding upper limit is slightly less constraining due to the weaker background reduction.

\section{Conclusions}

VERITAS observed the radio galaxy NGC~1275 (Perseus~A) for a total of 7.8~hours
good-quality live time between January and February 2009. The observation did not result in
a detection, despite the strong flux and hard spectrum recently reported by \emph{Fermi} up to 25~GeV
during the 2008 campaign and despite the similar flux during the VERITAS observation campaign.
Upper limits at 99\% confidence level are calculated using two analyses
optimized for Crab-like and softer spectra, respectively. In both cases, the flux upper limit
is incompatible with an extrapolation of the power law measured by \emph{Fermi}.
Excluding absorption by the EBL, due to the proximity of the galaxy,
the non-detection by VERITAS might be due to an intrinsic variability of the core. 
However, the \emph{Fermi}-LAT data during the VERITAS observation campaign
show a dim in flux by a factor 1.37 at the level of $2.8\,\sigma$ and no significant change in photon index,
compared to the 2008 data.
A simple variability of the flux is, therefore, not sufficient to explain the non detection by VERITAS.
The possibility of intrinsic spectral curvature at very high energy thus offers a viable explanation of the data,
but such a spectral variation must generate a strong drop of the flux at an energy not much above
the highest energy measured by \emph{Fermi}-LAT. The upper limit
in this work suggests an exponential  cutoff in the \emph{Fermi} spectrum at an energy $E\approx 18$~GeV, 
or $E\approx 120$~GeV in the case of a sub-exponential
cutoff. In either case the cutoff is likely above the maximum energy measured by \emph{Fermi} and both assumptions are compatible
with the \emph{Fermi} and VERITAS measurements.
The possibility that the source spectrum is a broken power law is also considered and is compatible with the data
with a break energy of about 16~GeV.

\acknowledgments

This research was supported by grants from the U.S. Department of
Energy, the U.S. National Science Foundation and the Smithsonian
Institution, by NSERC in Canada, by Science Foundation Ireland and by
STFC in the UK.



{\it Facilities:} \facility{FLWO (VERITAS)}.

\clearpage


\end{document}